\documentclass[12pt]{article}
 \usepackage{graphics}
\usepackage{amsmath}
\newcommand{\be}{\begin{equation}}
\newcommand{\ee}{\end{equation}}
\newcommand{\beq}{\begin{equation}}
\newcommand{\eeq}{\end{equation}}
\newcommand{\bea}{\begin{eqnarray}}
\newcommand{\eea}{\end{eqnarray}}

\begin{document}
\baselineskip=15.5pt \pagestyle{plain} \setcounter{page}{1}


\def\del{{\partial}}
\def\vev#1{\left\langle #1 \right\rangle}
\def\cn{{\cal N}}
\def\co{{\cal O}}
\newfont{\Bbb}{msbm10 scaled 1200}     
\newcommand{\mathbb}[1]{\mbox{\Bbb #1}}
\def\IC{{\mathbb C}}
\def\IR{{\mathbb R}}
\def\IZ{{\mathbb Z}}
\def\RP{{\bf RP}}
\def\CP{{\bf CP}}
\def\Poincare{{Poincar\'e }}
\def\tr{{\rm tr}}
\def\tp{{\tilde \Phi}}

\def\TL{\hfil$\displaystyle{##}$}
\def\TR{$\displaystyle{{}##}$\hfil}
\def\TC{\hfil$\displaystyle{##}$\hfil}
\def\TT{\hbox{##}}
\def\HLINE{\noalign{\vskip1\jot}\hline\noalign{\vskip1\jot}}
\def\seqalign#1#2{\vcenter{\openup1\jot
  \halign{\strut #1\cr #2 \cr}}}
\def\lbldef#1#2{\expandafter\gdef\csname #1\endcsname {#2}}
\def\eqn#1#2{\lbldef{#1}{(\ref{#1})}%
\begin{equation} #2 \label{#1} \end{equation}}
\def\eqalign#1{\vcenter{\openup1\jot
    \halign{\strut\span\TL & \span\TR\cr #1 \cr
   }}}
\def\eno#1{(\ref{#1})}
\def\href#1#2{#2}
\def\half{{1 \over 2}}

\def\ads{{\it AdS}}
\def\adsp{{\it AdS}$_{p+2}$}
\def\cft{{\it CFT}}

\newcommand{\ber}{\begin{eqnarray}}
\newcommand{\eer}{\end{eqnarray}}

\newcommand{\beqar}{\begin{eqnarray}}
\newcommand{\cN}{{\cal N}}
\newcommand{\cO}{{\cal O}}
\newcommand{\cA}{{\cal A}}
\newcommand{\cT}{{\cal T}}
\newcommand{\cF}{{\cal F}}
\newcommand{\cC}{{\cal C}}
\newcommand{\cR}{{\cal R}}
\newcommand{\cW}{{\cal W}}
\newcommand{\eeqar}{\end{eqnarray}}
\newcommand{\tht}{\thteta}
\newcommand{\lm}{\lambda}\newcommand{\Lm}{\Lambda}
\newcommand{\eps}{\epsilon}


\newcommand{\nonu}{\nonumber}
\newcommand{\oh}{\displaystyle{\frac{1}{2}}}
\newcommand{\dsl}
  {\kern.06em\hbox{\raise.15ex\hbox{$/$}\kern-.56em\hbox{$\partial$}}}
\newcommand{\id}{i\!\!\not\!\partial}
\newcommand{\as}{\not\!\! A}
\newcommand{\ps}{\not\! p}
\newcommand{\ks}{\not\! k}
\newcommand{\D}{{\cal{D}}}
\newcommand{\dv}{d^2x}
\newcommand{\Z}{{\cal Z}}
\newcommand{\N}{{\cal N}}
\newcommand{\Dsl}{\not\!\! D}
\newcommand{\Bsl}{\not\!\! B}
\newcommand{\Psl}{\not\!\! P}
\newcommand{\eeqarr}{\end{eqnarray}}
\newcommand{\ZZ}{{\rm \kern 0.275em Z \kern -0.92em Z}\;}


\def\azb{A_{\bar z}} \def\az{A_z} \def\bzb{B_{\bar z}} \def\bz{B_z}
\def\czb{C_{\bar z}} \def\cz{C_z} \def\dzb{D_{\bar z}} \def\dz{D_z}
\def\im{{\hbox{\rm Im}}} \def\mod{{\hbox{\rm mod}}} \def\tr{{\hbox{\rm Tr}}}
\def\ch{{\hbox{\rm ch}}} \def\imp{{\hbox{\sevenrm Im}}}
\def\trp{{\hbox{\sevenrm Tr}}} \def\vol{{\hbox{\rm Vol}}}
\def\rl{\Lambda_{\hbox{\sevenrm R}}} \def\wl{\Lambda_{\hbox{\sevenrm W}}}
\def\fc{{\cal F}_{k+\cox}} \def\vev{vacuum expectation value}
\def\nodiv{\mid{\hbox{\hskip-7.8pt/}}}
\def\ie{{\em i.e.}}
\def\ie{\hbox{\it i.e.}}

\def\CC{{\mathchoice
{\rm C\mkern-8mu\vrule height1.45ex depth-.05ex
width.05em\mkern9mu\kern-.05em} {\rm C\mkern-8mu\vrule
height1.45ex depth-.05ex width.05em\mkern9mu\kern-.05em} {\rm
C\mkern-8mu\vrule height1ex depth-.07ex
width.035em\mkern9mu\kern-.035em} {\rm C\mkern-8mu\vrule
height.65ex depth-.1ex width.025em\mkern8mu\kern-.025em}}}

\def\RR{{\rm I\kern-1.6pt {\rm R}}}
\def\NN{{\rm I\!N}}
\def\ZZ{{\rm Z}\kern-3.8pt {\rm Z} \kern2pt}
\def\IB{\relax{\rm I\kern-.18em B}}
\def\ID{\relax{\rm I\kern-.18em D}}
\def\II{\relax{\rm I\kern-.18em I}}
\def\IP{\relax{\rm I\kern-.18em P}}
\newcommand{\CS}{{\scriptstyle {\rm CS}}}
\newcommand{\CSs}{{\scriptscriptstyle {\rm CS}}}
\newcommand{\rc}{\nonumber\\}
\newcommand{\bear}{\begin{eqnarray}}
\newcommand{\eear}{\end{eqnarray}}
\newcommand{\W}{{\cal W}}
\newcommand{\F}{{\cal F}}
\newcommand{\x}{{\cal O}}
\newcommand{\LL}{{\cal L}}

\def\mani{{\cal M}}
\def\calo{{\cal O}}
\def\calb{{\cal B}}
\def\calw{{\cal W}}
\def\calz{{\cal Z}}
\def\cald{{\cal D}}
\def\calc{{\cal C}}
\def\to{\rightarrow}
\def\ele{{\hbox{\sevenrm L}}}
\def\ere{{\hbox{\sevenrm R}}}
\def\zb{{\bar z}}
\def\wb{{\bar w}}
\def\nodiv{\mid{\hbox{\hskip-7.8pt/}}}
\def\menos{\hbox{\hskip-2.9pt}}
\def\dr{\dot R_}
\def\drr{\dot r_}
\def\ds{\dot s_}
\def\da{\dot A_}
\def\dga{\dot \gamma_}
\def\ga{\gamma_}
\def\dal{\dot\alpha_}
\def\al{\alpha_}
\def\cl{{closed}}
\def\cls{{closing}}
\def\vev{vacuum expectation value}
\def\tr{{\rm Tr}}
\def\to{\rightarrow}
\def\too{\longrightarrow}

\newcommand{\gs}{\ensuremath{g_s}} 
\newcommand{\ap}{\ensuremath{\alpha'}} 
\newcommand{\ls}{\ensuremath{l_s}} 
\newcommand{\ms}{\ensuremath{m_s}} 
\newcommand{\lP}{\ensuremath{l_P}} 
\newcommand{\mP}{\ensuremath{m_P}} 

\def\eps{\epsilon}
\def\grad{\nabla}
\def\curl{\nabla\times}
\def\div{\nabla\cdot}
\def\p{\partial}
\def\pbar{\bar{\partial}}
\def\zbar{\bar{z}}
\def\para{{\scriptscriptstyle ||}}
\def\expec#1{\langle #1 \rangle}
\def\bra#1{| #1 \rangle}
\def\ket#1{\langle  #1 |}
\def\cbra#1{|\left. #1 \right)}
\newcommand{\Rea}{{\mathrm{Re}}}
\newcommand{\Ima}{{\mathrm{Im}}}
\newcommand{\cL}{{\mathcal{L}}}
\newcommand{\cZ}{{\mathcal{Z}}}
\newcommand{\bS}{{\mathbf{S}}}
\newcommand{\bT}{{\mathbf{T}}}
\newcommand{\bR}{{\mathbf{R}}}
\newcommand{\bZ}{{\mathbf{Z}}}
\newcommand{\cM}{{\mathcal{M}}}
\newcommand{\cG}{{\mathcal{G}}}

\newcommand{\tih}{\ensuremath{\tilde{h}}}
\newcommand{\Pt}{\ensuremath{P^{r}_{\theta}}}
\newcommand{\Px}{\ensuremath{P^{r}_{x}}}
\newcommand{\pix}{\ensuremath{\pi_{x}}}
\newcommand{\pit}{\ensuremath{\pi_{\theta}}}

%

\newfont{\namefont}{cmr10}
\newfont{\addfont}{cmti7 scaled 1440}
\newfont{\boldmathfont}{cmbx10}
\newfont{\headfontb}{cmbx10 scaled 1728}
 \numberwithin{equation}{section}

\begin{titlepage}

\begin{center} \Large \bf On Drag Forces and Jet Quenching in Strongly-Coupled Plasmas

\end{center}

\vskip 0.3truein
\begin{center}
Elena C\'aceres$^{\dagger}$\footnote{elenac@ucol.mx} and Alberto
G\"{u}ijosa$^{*}$ \footnote{alberto@nucleares.unam.mx}
\vspace{0.3in}

${}^{\dagger}$  Facultad de Ciencias,\\ Universidad de Colima,\\
Bernal D\'{\i}az del Castillo 340, Colima, Colima, M\'exico\\
 \vspace{0.3in}
$^{*}$ Departamento de F\'{\i}sica de Altas Energ\'{\i}as,
\\Instituto de Ciencias Nucleares, \\ Universidad Aut\'onoma de
M\'exico,\\ Apdo. Postal 70-543, M\'exico D.F.04510, M\'exico

\vspace{0.3in}

\end{center}
\vskip 1truein

{\bf ABSTRACT:} We compute the drag force experienced by a heavy
quark that moves through plasma in a gauge theory whose dual
description involves arbitrary metric and dilaton fields. As a
concrete application, we consider the cascading gauge theory at
temperatures high above the deconfining scale, where we obtain a
drag force with a non-trivial velocity dependence. We compare our
results with the jet-quenching parameter for the same theory, and
find qualitative agreement between the two approaches. Conversely,
we calculate the jet-quenching parameter for $\cN=4$
super-Yang-Mills with an R-charge density (or equivalently, a
chemical potential), and compare our result with the corresponding drag force.

\vspace{0.5in} \leftline{UTTG-08-06 } \leftline{ICN-UNAM-06/06G}
\vspace{0.2in}

\smallskip
\end{titlepage}
\setcounter{footnote}{0}

\section{Introduction and Summary}
Recently  there has been a surge of interest in the
possibility of employing the gauge/gravity duality
\cite{malda,gkpw} to determine the rate of energy loss in
finite-temperature strongly-coupled gauge theories. The ultimate
aim of this program would be to make contact with current
\cite{rhic} and future \cite{alice} experimental studies of
quark-gluon plasma (QGP), but at this point the gravity dual of
QCD is not yet available, so one must still be cautious when
attempting to draw inferences in this direction.

In the AdS/CFT context, the issue of energy loss has been
approached from three different perspectives. The authors of
\cite{Liu:2006ug} proposed a model-independent, non-perturbative
definition of the jet-quenching parameter (which in the QGP case
codifies the suppression of hadrons with high transverse momenta)
in terms of a light-like Wilson loop, which they then computed in
$\cN=4$ super-Yang-Mills (SYM) using the recipe of
\cite{maldawilson}. Their computation was generalized in
\cite{buchel} to the cascading gauge theory \cite{ks,kt} at finite
temperature \cite{ghkt}, and in \cite{vp} to certain marginal
deformations of the $\cN=4$ theory. Related work may be found in
\cite{Sin:2004yx, Shuryak:2005ia}.

A second approach was pursued in \cite{hkkky,gubser}, where the
drag force experienced by a heavy quark that moves through $\cN=4$
 SYM plasma was determined by considering a string
in the dual AdS-Schwarzschild geometry. This calculation was
extended to all asymptotically AdS geometries in \cite{herzog},
including the case dual to $\cN=4$ SYM with a non-zero chemical
potential, a case that was studied simultaneously in \cite{cacg}
from a different perspective: whereas \cite{cacg} worked directly
with the ten-dimensional spinning D3-brane background,
\cite{herzog} employed instead the five-dimensional charged black
hole solution of $\cN=8$ gauged supergravity \cite{Behrndt:1998jd}
obtained upon Kaluza-Klein reduction on the $S^5$
\cite{cveticgubser}, thereby arriving at different results. A more
detailed picture of the energy-loss process was painted in
\cite{Friess:2006aw}, which studied the wake left by the quark as
it ploughs through the plasma, using the methods of \cite{dkk,cg}.
The connection between the rate of energy loss found in
\cite{hkkky,gubser} and magnetic confinement was explored very
recently in \cite{Sin:2006yz}.

A third approach \cite{ct} extracted the diffusion coefficient for
a heavy quark in the $\cN=4$ plasma from an analysis of small
fluctuations of a Wilson line that follows the Schwinger-Keldysh
contour. On the AdS side of the duality, this involved a study of
 fluctuations that propagate along a string; similar calculations
 were carried out in \cite{hkkky}.

It is important to explore the relation between these three
approaches. As a step in this direction, in this paper we compare
the drag forces and jet-quenching parameters for two different
gauge theories. We begin in Section \ref{draggralsec} by
generalizing the drag force calculation of
\cite{hkkky,gubser,herzog} to backgrounds with arbitrary metric
and dilaton fields, finding the general result (\ref{dragforce}).
In Section \ref{dragktsec} we then specialize to the cascading
gauge theory, where the resulting drag force (\ref{dragKT}) is
found to display a highly non-trivial velocity dependence. We
compare this force with the jet-quenching parameter\footnote{To be
more precise, we find that the directly comparable quantities are
the jet-quenching parameter $\hat{q}$ and the ratio $\mu/T$, where
$\mu$ is the friction coefficient that determines the drag force
through $dp/dt=-\mu p$.} (\ref{qKT}) determined in \cite{buchel},
finding agreement in functional form, and numerical agreement (up
to an overall constant) in the region of large velocities. In
Section \ref{qrotsec} we proceed in the opposite direction,
computing the jet-quenching parameter for the $\cN=4$ plasma with
an R-charge density, and comparing the result with the drag force
determined in \cite{cacg}. This comparison is interesting because
in this case the quantities to be compared are not just numbers
but functions of the charge density $J$. For small values of $J$,
our result (\ref{qres}) is again in qualitative agreement with the
drag force (\ref{dragforcenlo}) obtained in \cite{cacg}, but for
arbitrary charges, the results (\ref{qrot}) and (\ref{dragrot})
disagree.

The general lesson appears to be that the parameter $\hat{q}$ defined by \cite{Liu:2006ug}
and the dissipative force extracted from the procedure pioneered in \cite{hkkky,gubser}
represent closely related but not identical measures of the rate of energy dissipation in a
given non-Abelian plasma. Our results underline the interest in exploring the connection
between the various approaches to energy loss from a more general viewpoint, by attempting
to extrapolate from one to the other directly at the level of the corresponding AdS/CFT
calculations.\footnote{After the first version of this paper had appeared on the arXiv, the
relation between the drag force \cite{hkkky,gubser} and jet-quenching \cite{Liu:2006ug}
approaches was discussed in the context of a study of mesons moving through the plasma
\cite{Liu:2006nn,cgg}. Related work may be found in \cite{qqbarrelated}.}

\section{Drag Force in Gauge Theories with Holographic \\ Duals}
\label{draggralsec}
 Consider a background dilaton field $\phi(x)$
and stationary Einstein frame metric
\begin{equation} \label{metric}
ds^2_E= G_{\mu\nu}(x) dx^\mu dx^\nu
\end{equation}
that holographically encode the dynamics of a strongly-coupled
gauge theory. Since we are interested in studying this gauge
theory at a finite temperature $T$, we assume the geometry
(\ref{metric}) includes a black hole \cite{wittenthermal}. In this
setup one can introduce an external quark in the gauge theory by
considering a string that has a single endpoint at the boundary
and extends all the way down to the horizon \cite{maldawilson}
(the gauge theory is therefore non-confining at the given
temperature).

The relevant string dynamics are captured by the Nambu-Goto action
\begin{equation} \label{NGaction}
S=-{1 \over 2 \pi \alpha'}\int  d\tau d\sigma e^{\phi/2}\sqrt
{-{\rm det}g_{\alpha\beta}}~,
\end{equation}
with $g_{\alpha\beta}= G_{\mu\nu}\del_\alpha X^\mu \del_\beta
X^\nu$. Letting $z$ denote the radial coordinate of the black hole
geometry and $t,x^i$ ($i=1,2,3$) label the directions along the
boundary at spatial infinity, we make the static gauge choice
$\sigma=z, \tau=t$, and, following \cite{hkkky,gubser},  focus
attention on the configuration
\begin{equation} \label{ansatz}
X^1(z,t) =vt +\rho(z)~,\quad X^2=0=X^3~.
\end{equation}
For the appropriate sign of $\rho'$, this describes the string
trailing behind its boundary endpoint as it moves at constant
speed $v$ in the $x^1$ direction, a configuration dual to the
external quark traversing the plasma.

Using (\ref{ansatz}) in (\ref{NGaction}), we find the Lagrangian
\begin{equation}
\cL\equiv e^{\phi/2}\sqrt{-{\rm det}g} = e^{\phi/2}\sqrt{
-G_{zz}G_{tt} -G_{zz}G_{xx} v^2 - G_{xx}G_{tt}\rho'^2}~,
\end{equation}
which results in an equation of motion for $\rho$ implying that
\begin{equation}
\pi_X = {\del \LL\over \del \rho'}= e^{\phi/2}{G_{xx}G_{tt} \over
\sqrt{-g}}\rho'
\end{equation}
is a constant. Inverting this relation we obtain
\begin{equation} \label{rhoprime}
(\rho')^2 =-\pi^2_\rho {{G_{zz}(G_{tt} +G_{xx}v^2)\over
G_{xx}G_{tt}(e^\phi G_{xx}G_{tt}+\pi_X^2 )}}~.
\end{equation}
Just like in the $\cN=4$ case analyzed in \cite{hkkky,gubser}, for
$v^2>0$ the numerator in this expression changes sign at a radius
$z=z_v$ defined by
\begin{equation} \label{zveq}
(G_{tt} +G_{xx}v^2)|_{z=z_v} =0~,
\end{equation}
and so the string will not be able to extend all the way down to
the horizon at $z=z_H$ unless the denominator also changes sign at
$z_v$.  This condition fixes
\begin{equation} \label{piXsol}
 \pi_X^2 = -e^\phi G_{xx}G_{tt}|_{z=z_v}=v^2 e^\phi G_{xx}^2|_{z=z_v}~.
 \end{equation}

With (\ref{rhoprime}) we can then compute the current density for
momentum along $x^1$,
\begin{eqnarray}
P^z_x&=&-{1\over 2\pi l_s^2}e^{\phi/2} G_{x\nu}g^{z\alpha}\del_\alpha X^\nu\\
&=& -{1\over 2\pi l_s^2}e^{\phi/2}{G_{xx}G_{tt}\over {\rm
det}g}\rho'~,
\end{eqnarray}
and use it to compute the drag force experienced by the
string/quark,
 \begin{equation}
 {dp_1 \over dt}={\sqrt{ -{\rm
det}g}}P_x^z = -{1\over 2\pi l_s^2}e^{\phi/2}{G_{xx}G_{tt}\over
\sqrt{ -{\rm det}g}}\rho'~,
\end{equation}
which after some algebra is easily seen to simplify to
\begin{equation}
{dp_1 \over dt}=-{\pi_X\over 2\pi l_s^2}=-{v\over 2\pi\ls^2}e^{\phi/2} G_{xx}|_{z=z_v}~.
\label{dragforce}
\end{equation}
This generalizes the result obtained in \cite{herzog} to
backgrounds with an arbitrary dilaton profile.

\section{Drag Force in the Cascading Plasma}
\label{dragktsec}

Let us now apply the results of the previous section to an
interesting concrete example: the Klebanov-Strassler cascading
gauge theory \cite{ks,kt}, whose dual geometry at temperatures
high above the deconfining transition was constructed in
\cite{ghkt} (see also \cite{buchelKT,bhkpt}) and is given by
 \begin{eqnarray} \label{kteq}
ds^2&=&{\sqrt{8a/K_*}\over\sqrt{z}}e^{2P^2\eta}\left(-(1-z)dt^2+d\vec{X}^2\right)+{\sqrt{K_*}\over
32}e^{-2P^2(\eta-5\xi)}{dz^2\over z^2(1-z)}
\\
{}&{}&\qquad\qquad+{\sqrt{K_*}\over 2}e^{-2P^2(\eta-\xi)}
\left[e^{-8P^2\omega}e^2_{\psi}
+e^{2P^2\omega}(e^2_{\theta_1}+e^2_{\phi_1}+e^2_{\theta_2}+e^2_{\phi_2})\right]~,
\nonumber \\
 \xi&=&{2z+[-2z+(z-2)\log(1-z)]\log z+(z-2){\rm Li}_2
(z)\over 40 K_* z}~, \nonumber \\
\eta&=&{ z-2 \over 16 K_* z} [\log z \log (1-z) +{\rm Li}_2 (z)]~,
\nonumber \\
 \phi&=& {P^2\over K_*}\left(-{\pi^2\over 24} +{1\over
4} {\rm Li_2}(1-z)\right)~, \nonumber
\end{eqnarray}
where the radial coordinate $z$ runs from the horizon at $z=1$ to
the boundary at $z\to 0$.

Using (\ref{kteq}) in  (\ref{piXsol})  one finds that $z_v=1-v^2$ and
\begin{equation} \label{Csoleq}
\pi_{X}=-v\left(e^{\phi/2}G_{xx}\right)_{z=z_v}=-\sqrt{8a\over
K_*}v\left(e^{2P^2\eta}e^{\phi/2}\over\sqrt{z}\right)_{z=1-v^2}~,
\end{equation}
or, more explicitly,
\begin{eqnarray} \label{CKTeq}
\pi_X&=&-\sqrt{8a\over K_*}{v\over\sqrt{1-v^2}}\exp\left\{
{P^2\over K_*}f(v)\right\}~,\\
f(v)&\equiv& {1\over 2}\left[-{\pi^2\over 24}+{1\over 4}{\rm Li}_2
v^2-{(1+v^2)\over 4(1-v^2)}\left(\log (1-v^2)\log v^2 +{\rm
Li}_2(1-v^2)\right)\right]~. \nonumber
\end{eqnarray}
The drag force is then given by (\ref{dragforce}) as
\begin{equation} \label{dragKT}
{dp\over dt}_{KT} = -{1\over 2\pi\ls^2}\sqrt{8a\over
K_*}{v\over\sqrt{1-v^2}} \left[1+ f(v){P^2\over K_*}
+\cO\left({P^4\over K^2_*}\right)\right]~,
\end{equation}
where we have kept only the first two terms in an expansion in
powers of $P^2/K_*\ll 1$, because the solution (\ref{kteq}) itself
is only valid to this order. The velocity-dependence seen in the
first term is just the $p/m$ factor present already in $\cN=4$ SYM
\cite{hkkky,gubser}. The second term has an additional non-trivial
dependence codified in the function $f(v)$ defined in
(\ref{CKTeq}), which approaches the value $-\pi^2/24\simeq -0.411$
for $v\to 0$, and has a logarithmic divergence in the
ultrarelativistic region $v\to 1$. As seen in Fig.~1, away from
this region $f(v)$ is nearly constant.

\begin{figure}
\centerline{\includegraphics{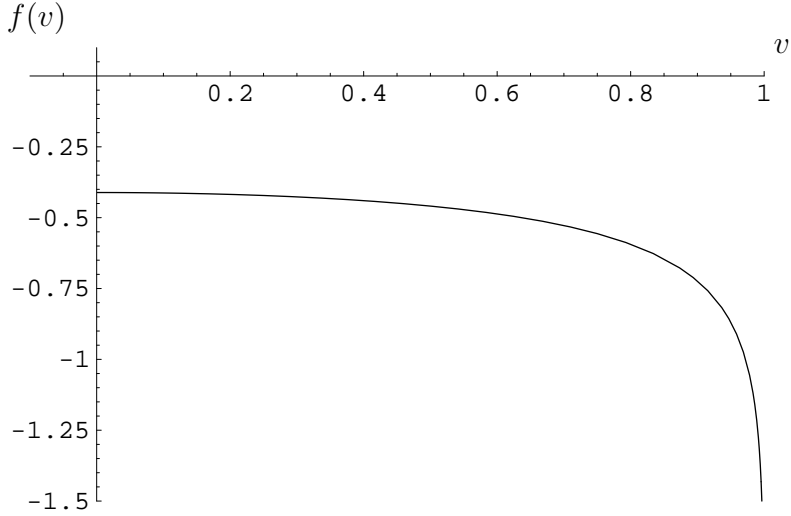}} \label{figv}
 \begin{picture}(0,0)
  \put(90,200){$f(v)$}
  \put(380,190){$v$}
   \end{picture}
 \vspace*{-0.5cm}\caption{\small The
function defined in (\ref{CKTeq}), which codifies the dependence
of the drag force (\ref{dragKT}) on the velocity beyond the
expected relativistic dependence $v/\sqrt{1-v^2}$ found already in
$\cN=4$ SYM.} \label{figv}
\end{figure}

As in \cite{buchel}, it is instructive to compare the result
(\ref{dragKT}) for the cascading plasma against the drag force in
$\cN=4$ SYM \cite{hkkky,gubser},
\begin{equation}
 { dp \over dt}_{\cN=4} = - \frac{\pi \sqrt{g^2_{YM}N}}{2}
T^2 {v \over \sqrt{1-v^2}}~,
\end{equation}
by considering the ratio
\begin{equation}
{{dp\over dt}_{KT} \over { dp \over dt}_{\cN=4}}=
{1\over\pi^2\ls^2}\sqrt{8a\over K_*}{1\over T^2}{1\over
\sqrt{g^2_{YM}N}}\left[1+ {P^2\over K_*} f(v) +\cO\left({P^4\over
K^2_*}\right)\right]~.
\end{equation}
Employing the relation $8a= T^4 K^2_*/4$ \cite{ghkt}, this reads
\begin{equation} {{dp\over
dt}_{KT} \over { dp \over dt}_{\cN=4}}= {1\over
2\pi^2\ls^2}\sqrt{{K_*\over g^2_{YM}N} } \left[1+ {P^2\over K_*}
f(v) +\cO\left({P^4\over K^2_*}\right)\right]~.
\end{equation}
Upon synchronizing the rank of the $\cN=4$ theory with the
effective rank of the cascading gauge theory at the given
temperature by setting $K_*=\sqrt{2/\pi}L_P^4
N=\sqrt{4\kappa^2\pi}N=2^4\pi^3\gs\ls^4 N$, and using
$g_{YM}^2=4\pi\gs$, we are left with the final result
\begin{equation} \label{dragratio}
{{dp\over dt}_{KT} \over { dp \over dt}_{\cN=4}}= 1+ f(v){P^2\over
K_*}  +\cO\left({P^4\over K^2_*}\right)~.
\end{equation}
This ratio has the same qualitative form as the one computed for
the corresponding jet-quenching parameter calculated  in
\cite{buchel},
\begin{equation} \label{qKT}
{{\hat q}_{cascade}\over {\hat q}_{{\cal N}=4}} =1 +\chi {P^2\over
K_*}  +\cO\left({P^4\over K^2_*}\right)~,
\end{equation}
with $\chi\simeq -1.388$. In both cases the temperature-dependence
arises only from the dependence of the effective rank $K_*$ on
$T$, the precise form of which is
 $K_*/2P^2\simeq\log(T/\Lambda)$ \cite{ghkt}. In addition, $f(v)$ is
negative for any value of $v$ (see Fig.~\ref{figv}), which implies
that, just like the jet-quenching parameter, the drag force
increases with increasing temperature. It also seems worth
pointing out that precise numerical agreement between $f(v)$ and
$\chi$ is achieved at a rather large value of the velocity, $v\sim
.994$, which appears to be related to the fact that the
jet-quenching calculation focuses on ultra-relativistic quarks.

Before closing this section we should note that it has recently been argued
\cite{PandoZayas:2006sa} that (\ref{kteq}) cannot be trusted all the way down to the
boundary at $z=0$, because the perturbative expansion in $P^2/K_*\ll 1$ through which it was
derived \cite{ghkt} breaks down at a critical radius $z_c=e^{-2K_*/P^2}$, which is
non-perturbatively small but non-zero. Since the drag force (\ref{dragKT}) is determined by
the value of the metric and dilaton at $z_v=1-v^2$, this does not affect our result as long
as $z_v\gg z_c$, which means that we can consider ultra-relativistic velocities except for a
region parametrically close to $v=1$, defined by the condition $\ln\gamma\ge K_*/P^2$. As an
example, validity of (\ref{dragKT}) at the velocity $v=0.994$ considered in the preceding
paragraph merely requires that $P^2/K_*\ll 0.45$. The problem of obtaining a solution valid
all the way down to $z=0$ has been addressed numerically in \cite{PandoZayas:2006sa}. In any
case, our main goal in this section has been to obtain a drag force comparable to the
jet-quenching parameter (\ref{qKT}), which was derived in \cite{buchel} using the background
(\ref{kteq}).

\section{Jet-Quenching Parameter in a Charged $\cN=4$ Plasma}
\label{qrotsec}

The near-horizon metric for rotating D3-branes at finite
temperature and with one angular momentum turned on is
\cite{gubserspinning,russo,klt}
\begin{eqnarray} \label{metric}
ds^2&=&{1\over\sqrt{H}}\left[{(1-h)\over 2}((dx^+)^2
+ (dx^-)^2) -(1+h)dx^+ dx^- + dx_2^2 + dx_3^2 \right]+\\
& &\quad \quad \quad\sqrt{H}\left[{dr^2\over\tilde{h}} -{2lr_0^2 \over R^2}\sin^2\theta
dtd\phi +r^2(\Delta d\theta^2 +\tilde{\Delta}\sin^2\theta d\phi^2 + \cos^2\theta
d\Omega_3^2)\right], \nonumber
\end{eqnarray}
where
\begin{eqnarray}
x^{\pm}&=&{1\over\sqrt{2}} (x^0\pm x^1)~,\nonumber\\
H&=&\frac{R^4}{r^4\Delta}~,\nonumber\\
\Delta&=&1+\frac{l^2\cos^2\theta}{r^2}~,\nonumber\\
\tilde{\Delta}&=&1+{l^2\over r^2}~, \nonumber\\
h&=&1-{r_0^4\over r^4\Delta}~, \nonumber\\
\tilde{h}&=&{1\over\Delta}\left(1+{l^2\over r^2}-{r_0^4\over
r^4}\right)~,\nonumber\\
R^4&=& 4\pi N\gs\ls^4~. \label{L}
\end{eqnarray}
This background is dual to $\cN=4$ SYM theory  with an R-charge
density, or equivalently, a chemical potential. The geometry has
an event horizon at the positive root of $\tilde{h}(r_H)=0$,
\begin{equation} \label{rH}
r_H^2={1\over 2}\left(\sqrt{l^4+4r_0^4}-l^2\right)~,
\end{equation}
and an associated Hawking temperature, angular momentum density, and angular velocity at the
horizon
\begin{equation} \label{Tj}
T={r_H\over {2\pi R^2 r_{0}^2}}\sqrt{l^4+4r_0^4}~,\qquad J={l r_0^2 R^2\over 64
\pi^4\gs^2\ls^8}~,\qquad \Omega={l r_H^2\over r_0^2 R^2}~,
\end{equation}
which  translate respectively into the temperature, R-charge density and R-charge chemical
potential in the gauge theory.

 In this section we will calculate the jet-quenching parameter
$\hat{q}_J$ following \cite{Liu:2006ug}, and compare with the drag force result of
\cite{cacg}. For this we must consider a string whose endpoints lie at $r\to\infty$ in the
spacetime (\ref{metric}) and trace a rectangular light-like Wilson loop of length $L$ along
$x^2\equiv y$ and $L^-$ along $x^-$ \cite{Liu:2006ug} . Making the static gauge choice
$\sigma=y, \tau=x^-$, the relevant configuration is $r(y,x^-)=r(y)$, with all other
embedding fields constant, and with boundary conditions $r(\pm L/2)=\infty$. The Nambu-Goto
action reduces to
\begin{equation} \label{NGaction}
S= {{\sqrt 2} L^-\over 2\pi\alpha'}\int ^{L/2}_{0
}dy\,\sqrt{\left({1\over H} + {(r')^2 \over \tilde
h}\right)(1-h)}~.
\end{equation}
Notice that, just like in \cite{Liu:2006ug,buchel,vp}, we are
working in Lorentzian signature and have omitted a factor of $i$
in (\ref{NGaction}).

Regarding $y$ as `time', the fact that the Lagrangian is
time-independent implies that the `energy' is conserved, a
statement that is easily seen to lead to
\begin{equation} \label{rprime}
(r')^2= {\tilde h\over H}\left({\gamma (1-h)\over H}
-1\right)={\tilde h\over H}\left({\gamma r_0^4\over R^4}
-1\right)~,
\end{equation}
with $\gamma\ge R^4/r_0^4$  an integration constant. It follows
from this equation that the minimum value of $r(y)$, which by
symmetry must lie at $y=0$, occurs at the radius where
$\tilde{h}=0$, i.e., at the horizon $r_H$.

Integrating (\ref{rprime}) we find a relation between $\gamma$ and
the width $L$ of the Wilson loop,
\begin{equation} \label{L}
{L\over 2 }= {1\over \sqrt{(\gamma r_0^4/R^4 -1)}}\int
_{r_H}^{\infty} dr \sqrt{H\over \tilde h}= {R^2 I\over
r_H\sqrt{(\gamma r_0^4/R^4 -1)}}~,
\end{equation}
where we have defined
\begin{equation} \label{I}
I=\int _{1}^{\infty}  {d\rho\over \sqrt{\rho^4+(l^2/r_H^2)
\rho^2-r_0^4/r_H^4}}=\int _{0}^1 { d\zeta \over   \sqrt{1
+(l^2/r_H^2)\zeta ^2  -r_0^4/r_H^4 \zeta ^4}}~.
\end{equation}
 Also, using (\ref{rprime}) in (\ref{NGaction}) we
are left with a trivial integral that yields
\begin{equation} \label{S}
S={L^-L\over 2\sqrt {2} \pi\alpha '}\sqrt{\gamma}~.
\end{equation}
According to the recipe of \cite{maldawilson}, to compute the
Wilson loop we must subtract from $S$ the self-interaction of the
isolated quark and the isolated antiquark, which in the AdS side
corresponds to the Nambu-Goto action evaluated for strings that
extend from (to) $r\to\infty$ to (from) $r=r_H$ at fixed $y$,
\begin{equation} \label{s0}
S_0= {\sqrt {2} L^-\over 2\pi\alpha'}\int ^{\infty}_{r_h}dr
\sqrt{1-h\over \tilde h}={\sqrt {2} r_0^2 L^- I\over 2\pi\alpha'
r_H}~.
\end{equation}

It is clear from (\ref{L}) that small $L$ corresponds to large
$\gamma$. In this regime, the leading term in (\ref{L}) implies
$\sqrt{\gamma}\propto 1/L$, which when substituted in (\ref{S})
gives an $L$-independent result that is precisely cancelled by
(\ref{s0}). The quantity we are after comes from the
next-to-leading term in (\ref{L}) for large $\gamma$, which yields
\begin{equation}
S_I=S-S_0={ r_0^2 r_H L^- L^2\over 8\sqrt {2}\pi\alpha' R^4 I}~.
\end{equation}
Using the definition of \cite{Liu:2006ug}, the jet-quenching parameter then follows
as\footnote{We employ here the final normalization of $\hat{q}$ given in v3 of
\cite{Liu:2006ug}; the original definition was a factor of $\sqrt{2}$ smaller.}
\begin{equation} \label{qrot}
\hat{q}_J\equiv {2S_I\over L^- L^2/4\sqrt{2}}={r_0^2 r_H\over \pi\ap R^4 I}~.
\end{equation}

The final step would be to express the result (\ref{qrot}) in terms
of gauge theory quantities. This can be done analytically in the
$l\ll r_0$ regime, where up to $\cO(l^4/r_0^4)$ corrections the
relations (\ref{rH}) and (\ref{Tj}) imply
\begin{equation}
r_H\simeq r_0\left(1-{l^2\over 4
r_0^2}\right)=r_0\left(1-{4J^2\over \pi^2 N^4 T^6}\right), \qquad
r_0\simeq \pi R^2 T\left(1+{4J^2\over\pi^2 N^4 T^6}\right)~,
\end{equation}
and the integral (\ref{I}) can be seen to give
\begin{equation}
I\simeq {\sqrt{\pi}\Gamma(5/4)\over \Gamma(3/4)} + {l^2 \over 4
r_0^2}( \mathrm{E}(-1)- 2\mathrm{K}(-1))\equiv a - {l^2 \over 4
r_0^2}b~,
\end{equation}
where E and K respectively denote complete elliptic integrals of
the second and first kind. The numerical value of the coefficients
defined above is $a\simeq 1.311, b\simeq  0.7120$. Putting this
all together, and remembering that the 't~Hooft coupling  is given
by $\lambda\equiv g_{YM}^2 N =R^4/\ap$, we are finally left with
\begin{equation} \label{qres}
\hat{q}_J={\pi^2\sqrt{\lambda}T^3\over a}\left[1+ {4 J^2\over \pi^2 N^4 T^6}(2+b/a) + {\cO}(
{J^4\over T^{12}}) \right]~.
\end{equation}
As a (rather mild) check, note that the leading term,
$\hat{q}_{J=0}$, reproduces the result of \cite{Liu:2006ug}. The
next-to-leading term is a new result.

Again, it is natural to compare (\ref{qres}) against the drag
force calculated in \cite{cacg} (see also \cite{herzog}),
\begin{equation} \label{dragforcenlo}
\left({dp_1 \over dt}\right)_J=-{\pi\over 2}{p_1\over
m}{\sqrt{\lambda} T^2}\left[1+{8 J^2\over \pi^2 N^4 T^6} +
\cO(J^4/T^{12})\right]~.
\end{equation}
The comparison is especially interesting because the quantities to
be compared are now \emph{functions} of the additional parameter
$J$. The leading term in (\ref{dragforcenlo}) is of course the
result computed in \cite{hkkky,gubser} at zero chemical potential.
As in the cascading gauge theory case analyzed in the previous
section, we find that the subleading terms in the jet-quenching
ratio

\begin{equation}
 { {\hat q}_J \over {\hat q}_{0}}=1+ {8 J^2\over
\pi^2 N^4 T^6}(1+b/2a) + {\cO}( {J^4\over T^{12}})
\end{equation}
and the drag force ratio,
\begin{equation}
{ (dp_1/dt)_J \over (dp_1/dt)_{0}}=1+{8 J^2\over \pi^2 N^4 T^6} +
\cO(J^4/T^{12})
\end{equation}
have the same qualitative form.The numerical
coefficients are also in rough agreement: they are both positive,
implying that both the drag force and the jet-quenching parameter
increase with increasing charge density, and they are of the same
order of magnitude.

It is important to note, however, that this agreement cannot
persist at arbitrarily high order in the expansion in powers of
$l/r_0$, because the full result (\ref{qrot}) for the
jet-quenching parameter evidently has a different functional
dependence on $r_0$ and $l$ than the full drag force result
\cite{cacg}
\begin{equation} \label{dragrot}
\left({dp_1\over dt}\right)_J=-{{r_0^2/R^2}\over
2\pi\ls^2}{p_1\over m}~.
\end{equation}
This appears to imply that, in the general case, the drag force
and jet-quenching parameter codify somewhat different information
on the process of energy loss in a plasma.

After the first version of this paper had been posted on the arXiv, three other calculations
of $\hat{q}$ in a charged $\N=4$ SYM plasma appeared
\cite{Lin:2006au,Avramis:2006ip,Armesto:2006zv}. The first of these is not directly
comparable to ours, because the authors of \cite{Lin:2006au} employed a five-dimensional
supergravity perspective instead of the ten-dimensional string theory viewpoint adopted here
(for a discussion on the relation between these two approaches, see \cite{cacg}). Our full
result (\ref{qrot}) for the singly-charged plasma agrees with the one obtained by the
authors of \cite{Avramis:2006ip} (who also determined $\hat{q}$ for two equal non-zero
charges) and \cite{Armesto:2006zv} (who addressed the general three-charge case). This
result was plotted in \cite{Avramis:2006ip} and shown to be non-monotonic beyond the
small-charge region explored in (\ref{qres}), which motivated the authors of \cite{cacg} to
generate a comparable plot of their drag force result (\ref{dragrot}). As expected from the
discussion above, the two graphs are different but qualitatively similar, so again the
general lesson seems to be that the parameter $\hat{q}$ defined by \cite{Liu:2006ug} and the
dissipative force extracted from the procedure pioneered in \cite{hkkky,gubser} represent
closely related but not identical measures of energy dissipation in a given non-Abelian
plasma. Evidently, more work will be required to completely elucidate the relation between
these two approaches.

\section{Acknowledgments}
It is a pleasure to thank Mariano Chernicoff and Jos\'e~Antonio
Garc\'{\i}a for useful conversations. Elena C\'aceres thanks the
Theory Group at the University of Texas at Austin for hospitality
during the completion of this work. Her research is supported in
part by the National Science Foundation under Grant No.
PHY-0071512 and PHY-0455649;  by Mexico's
National Council for Science and Technology under grant CONACYT No50760 and by the University of Colima under grant FRABA  No.447/06.
The research of Alberto G\"uijosa is supported in part by Mexico's
National Council for Science and Technology grants CONACyT
40754-F, CONACyT SEP-2004-C01-47211 and CONACyT/NSF J200.315/2004,
as well as by DGAPA-UNAM grant IN104503-3.

\end{document}